\documentclass{article}

\usepackage[utf8]{inputenc}
\usepackage[T1]{fontenc}
\usepackage{amsmath,amssymb}
\usepackage{graphicx}
\usepackage{caption}
\usepackage{times}
\usepackage{lineno}

\usepackage{color}
\usepackage{authblk}

\title{Uni-Traveling-Carrier Photodiode Based on MoS\(_2\)/GaN van der Waals Heterojunction for High-Speed Visible-Light Detection}
\date{}

\author[1]{Takuya Kadowaki}
\author[1]{Takahiro Serikawa}
\author[1]{Akihide Ichikawa}
\author[1]{Yuji Ohmaki}
\author[1]{Koji Usami}
\author[1]{Yoichi Kawakami}
\author[2]{Yoshihiro Iwasa}
\author[1]{Hisashi Ogawa}

\affil[1]{Nichia Corporation, Yokohama, Kanagawa 221-0022, Japan}
\affil[2]{RIKEN Center for Emergent Matter Science (CEMS), Wako, Saitama 351-0198, Japan}

\begin{document}

\maketitle

\begin{abstract}
Uni-traveling-carrier photodiodes (UTC-PDs), which utilize only electrons as the active carriers, have become indispensable in high-speed optoelectronics due to their unique capabilities, such as high saturation power and broad bandwidth. 
However, extending the operating wavelengths into the visible region for wider applications is challenging due to the lack of suitable wide-bandgap III-V semiconductor combinations with the necessary band alignment and lattice matching. 
Here, we show that a UTC-PD based on a van der Waals heterojunction composed of a 2D transition metal dichalcogenide, molybdenum disulfide (MoS\(_2)\), as a photoabsorption layer and a gallium nitride (GaN) film as a carrier transport layer, offers a solution to this challenge. 
The fast vertical carrier transport across the heterointerface is enabled by the direct epitaxial growth of a MoS\(_2\) layer on a GaN film. 
Our device demonstrates a frequency response in the several-GHz range with a quantum efficiency on the order of 1\% throughout the entire visible spectrum, highlighting the promise for high-speed visible optoelectronics.

\end{abstract}

\section*{} \label{sec:1}

Wide-bandgap semiconductors offer intrinsically favorable electronic properties such as high-power operation, high breakdown voltage, and high-temperature stability \cite{iannaccone2021power,Baliga_2013}. 
In particular, gallium nitride (GaN) has attracted attention owing to its high electron mobility, large electron saturation velocity \cite{electronics8050575,carey2019extreme}. 
GaN-based transistors have already become key elements in high-frequency electronic devices and wireless communications \cite{8962057,LU2025315}.
It is thus natural to anticipate that GaN-based photodiodes can exhibit excellent high-speed high-power characteristics in the visible wavelength range.
Such high-speed wide-bandgap photodiodes can become key components in emerging visible-light technologies such as intersatellite communication \cite{8353824}, visible-light communication (Li-Fi) including underwater communication \cite{Qi:23,cryst11091098}, and high-resolution distance sensing technologies such as the Light Detection And Ranging (LiDAR) and Optical Coherence Tomography (OCT) \cite{Mitchell_Hoehler_Giorgetta_Hayden_Rieker_Newbury_Baumann_2018,Xu_Zhang_Abbasi_Liu_Yan_Ng_Yuan_2024}.

Uni-traveling carrier photodiodes (UTC-PDs) \cite{Ishibashi:97,Ishibashi_Ito_2020} are the most successful example of high-speed photonic devices in the telecommunication industry.
They utilize the type-II band-alignment of the photoabsorption layer and diffusion block layer, which allow only electrons to transit in the carrier transport layer, to maximize response bandwidth and suppress space-charge saturation \cite{app14083410}. 
Adapting this concept to the visible range is nonetheless challenging, as it requires semiconductor combinations with specific band alignment and good lattice matching \cite{Piels:12}, which, to our knowledge, have not yet been realized for GaN-based materials. 
Moreover, InGaN ternary alloys need to be inserted to shift the absorption edge of GaN (360 nm, corresponding to the bandgap \(\sim\)3.4 eV) to the visible range. 
The InGaN/GaN quantum-well structure, however, reduces bandwidth due to slow carrier escape dynamics, resulting in the bandwidth typically in several hundred MHz to a few GHz \cite{Hu_Chen_Li_Zou_Zhang_Hu_Zhang_He_Yu_Jiang_2021, Xu_Luo_Lin_Shen_Wang_Zhang_Wang_Jiang_Chi_2023}. 

Recent advances have shown that two-dimensional (2D) metal dichalcogenides, such as molybdenum disulfide (MoS\(_2\)), offer unique electrical and photonic properties.
Combined with GaN, MoS\(_2\) can become a high-performance photoabsorption layer since MoS\(_2\) shows the small lattice mismatch \cite{LayeredTMD2018,Yan_Tian_Yang_Weng_Zhang_Wang_Xie_Lu_2018,ALKHALFIOUI2025128047}, type-II band alignment \cite{Tangi_Mishra_Ng_Hedhili_Janjua_Alias_Anjum_Tseng_Shi_Joyce_Li_Ooi_2016,Moun2018,FirstPrinciples2019} and strong light-matter interaction in the visible region \cite{Mak2016,Taffelli2021,Jian_Cai_Zhao_Li_Zhang_Liu_Xu_Liang_Zhou_Dai_2023}. 
Although several groups explored MoS\(_2\)/GaN heterojunction photodetectors \cite{Moun2020,Jain2020,Veeralingam2023,Janardhanam_Zummukhozol_Jyothi_Shim_Choi_2023,Vashishtha2024}, relatively slow responses on the order of microseconds or longer \cite{Zhuo2019,doi:10.1021/acsami.0c11021,PulsedLaserDeposition2023,Ye_Gan_Schirhagl_2024} have been reported. 
This is significantly slower than expected, contradicting the ultrafast (\(\sim\)20 fs) carrier transport dynamics in 2D thin layers reported \cite{Jin2018,Li2022}. 
Such slow responses are often attributed to the imperfect interface of the heterojunction \cite{PulsedLaserDeposition2023}, and the intrinsic speed capabilities of MoS\(_2\)/GaN interfaces have remained unclear.

In this paper, we report the successful realization of a UTC-PD based on a MoS\(_2\)/GaN van der Waals heterojunction, which exhibits frequency response exceeding several GHz for visible light. 
Here, a few-layer MoS\(_2\) absorption layer is epitaxially grown directly on a GaN film using metal-organic chemical vapor deposition (MOCVD), as opposed to the stamp transfer method.
The demonstrated response of our device is currently constrained by external electronic cutoff rather than fundamental carrier transport limits, suggesting that further structural optimization can push the bandwidth to tens of GHz or higher.


\section*{Formation of MoS\(_2\)/GaN Type-II Heterojunction} \label{sec:2}

The structure and operation principle of the MoS\(_2\)/GaN UTC-PD developed in this study is schematically illustrated in Fig.~\ref{fig:MoS${}_2$GaN_overview}\textbf{a}.
The device is composed of an absorption layer of MoS\(_2\), an i-GaN carrier transport layer, and an n-GaN contact layer. 
Incident photons absorbed within the MoS\(_2\) layer generate electron-hole pairs.
The holes rapidly diffuse to the anode electrode adjacent to the MoS\(_2\) layer, whereas the electrons traverse the type-II aligned conduction band interface into the i-GaN layer.
Subsequently, the electrons alone travel through the externally biased i-GaN layer and eventually reach the cathode.
Similar to standard UTC-PDs, high-speed operation is anticipated because of the smaller effective mass and the larger saturation velocity of electrons (\(\sim 3\times 10^7\) cm/s) than those of holes (\(\sim 0.6\times 10^7\) cm/s) \cite{10.1063/1.2828003}.

To achieve optimal interface conditions, we employed epitaxial growth of the MoS\(_2\) layers directly onto a GaN film on a c-plane sapphire substrate using metal-organic chemical vapor deposition (MOCVD). 
More detailed information on the growth conditions, are described in the Method section.

Raman spectroscopy analysis reveals a frequency difference \(\Delta\omega\) of 22 cm\(^{-1}\) between the E\(^{1}_{2g}\) and A\(_{1g}\) peaks (Fig.~\ref{fig:MoS${}_2$GaN_overview}\textbf{b}), indicating that the MoS\(_2\) layers consist of bilayer or trilayer \cite{ChangguLee2010}. 
The atomic alignment of MoS\(_2\) layers on GaN is confirmed by cross-sectional high-resolution scanning transmission electron microscopy (STEM), shown in Fig.~\ref{fig:MoS${}_2$GaN_overview}\textbf{c}.
Remarkably, the MOCVD-deposited MoS\(_2\) layers are epitaxially grown on the GaN film, likely due to the small lattice mismatch (0.8\%) between MoS\(_2\) and GaN crystals \cite{LayeredTMD2018}.
The absence of unintended interfacial layers, such as native oxide, is also evidenced by the small lattice mismatch between measured and theoretical interlayer atomic spacing (Extended Data Fig. 1).

Figure~\ref{fig:MoS${}_2$GaN_overview}\textbf{d} shows the photoabsorption properties of the MoS\(_2\)/GaN heterostructure.
Absorption peaks are attributed to the excitonic peaks A, B, and C of the bare MoS\(_2\) absorption \cite{Photocarrier2014,Wang2017} (inset in Fig.~\ref{fig:MoS${}_2$GaN_overview}\textbf{d}).

In order to examine the band alignment of the MoS\(_2\)/GaN van der Waals heterostructure, Kelvin Probe Force Microscopy (KPFM) measurements were conducted. 
This technique is capable of probing the surface potential of materials, making it a widely used method for determining the band offsets in heterostructures involving 2D materials~\cite{Mahmut2015,Giannazzo2020,Jiang2024}. 
Figure~\ref{fig:KPFM_SurfacePotential}\textbf{a} shows the mapping image of the MoS\(_2\) on GaN/sapphire substrate, showing change in surface potential between MoS\(_2\) and GaN, 
and a line-scan of the surface potential along the indicated line in this image is reported in Fig.~\ref{fig:KPFM_SurfacePotential}\textbf{b}.
The results indicate the type-II band alignment of the conduction band and the offset \(\Delta E_{c}\) of 0.26 eV, as shown in Fig.~\ref{fig:KPFM_SurfacePotential}\textbf{c}, which is consistent with previous reports~\cite{Moun2018,BandAlignment2013,FirstPrinciples2019}.

These results verify that high-quality MoS\(_2\) layers effectively function as the active absorption region for UTC-PD applications.

\begin{figure}[htbp]
    \centering
\includegraphics[width=\textwidth]{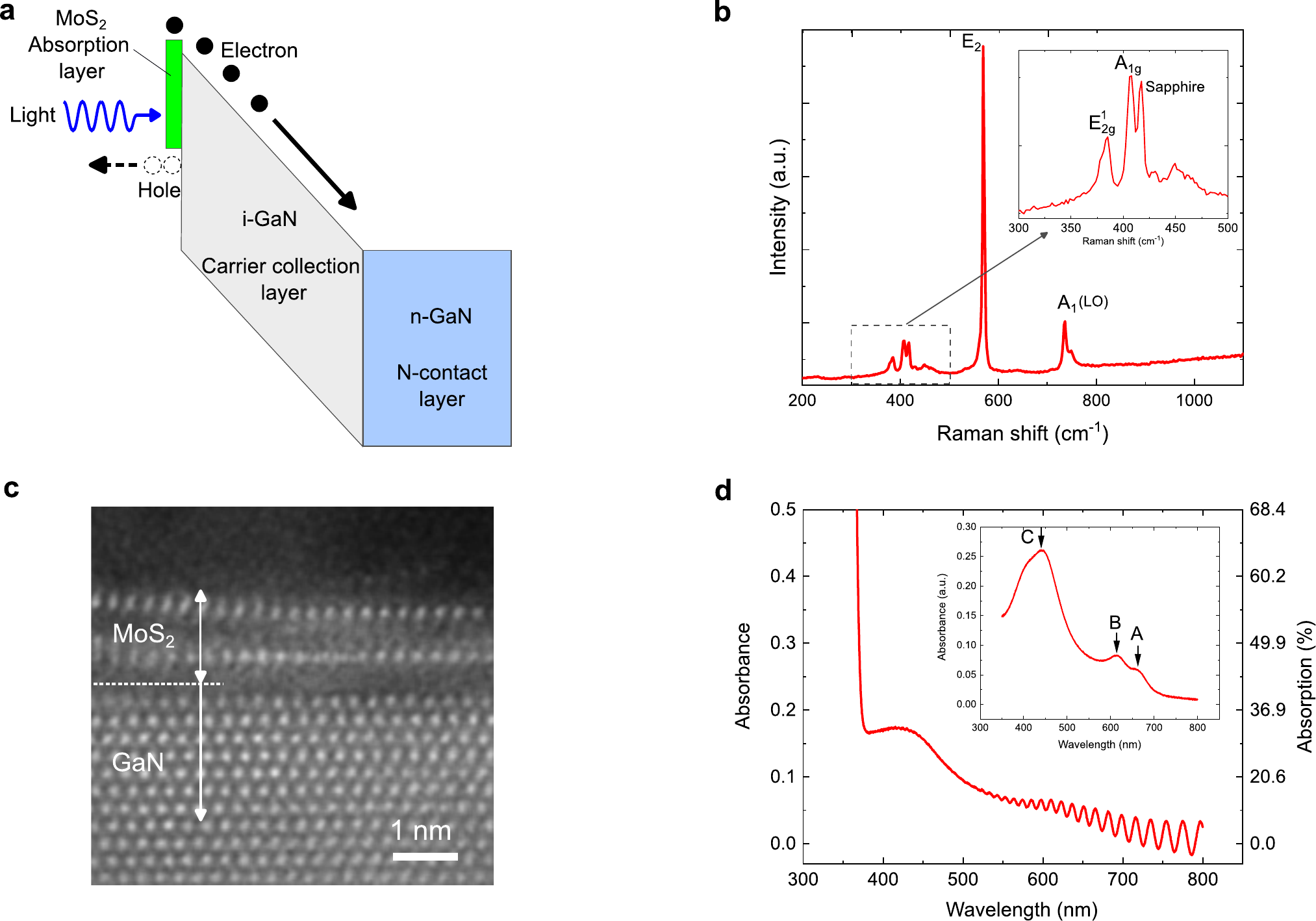}
    \caption{\textbf{a}, Band structure of a MoS\(_2\)/GaN UTC-PD. \textbf{b}, Raman spectra of the MoS\(_2\)layer grown on a GaN/sapphire substrate. The inset shows the details of the Raman peaks due to the MoS\(_2\) layer. \textbf{c}, High resolution high-angle annular dark field (HAADF) cross-sectional STEM image of MoS\(_2\)/GaN heterojunction. The MoS\(_2\) film consists of 2 to 4 layers, being consistent with the result of the Raman spectroscopy. \textbf{d}, Absorption spectrum of the MoS\(_2\)/GaN heterostructure measured by a spectrophotometer. At wavelengths below 380 nm, absorption predominantly originates from the GaN layer. For the reference, the inset graph shows the absorption spectrum of the bare MoS\(_2\) layer directly grown on the sapphire substrate.}
    \label{fig:MoS${}_2$GaN_overview}
\end{figure}

\begin{figure}[htbp]
    \centering
\includegraphics[width=\textwidth]{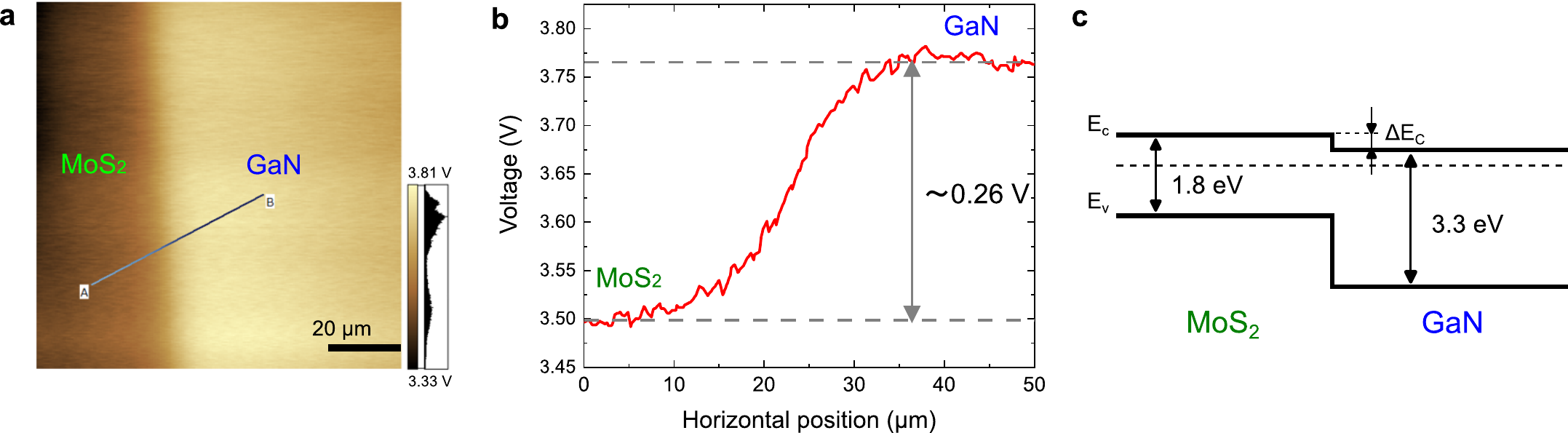}
    \caption{\textbf{a}, Kelvin Probe Force Microscope (KPFM) mapping image of MoS\(_2\) on GaN/sapphire substrate, showing change in surface potential between MoS\(_2\) and GaN. \textbf{b}, Plot of surface potential difference across MoS\(_2\)/GaN interface along the line as indicated in (\textbf{a}). \textbf{c}, Schematic of the band alignment of the MoS\(_2\)/GaN van der Waals heterojunction.}
    \label{fig:KPFM_SurfacePotential}
\end{figure}

\section*{Device Fabrication} \label{sec:3}

\begin{figure}[htbp]
    \centering
\includegraphics[width=\textwidth]{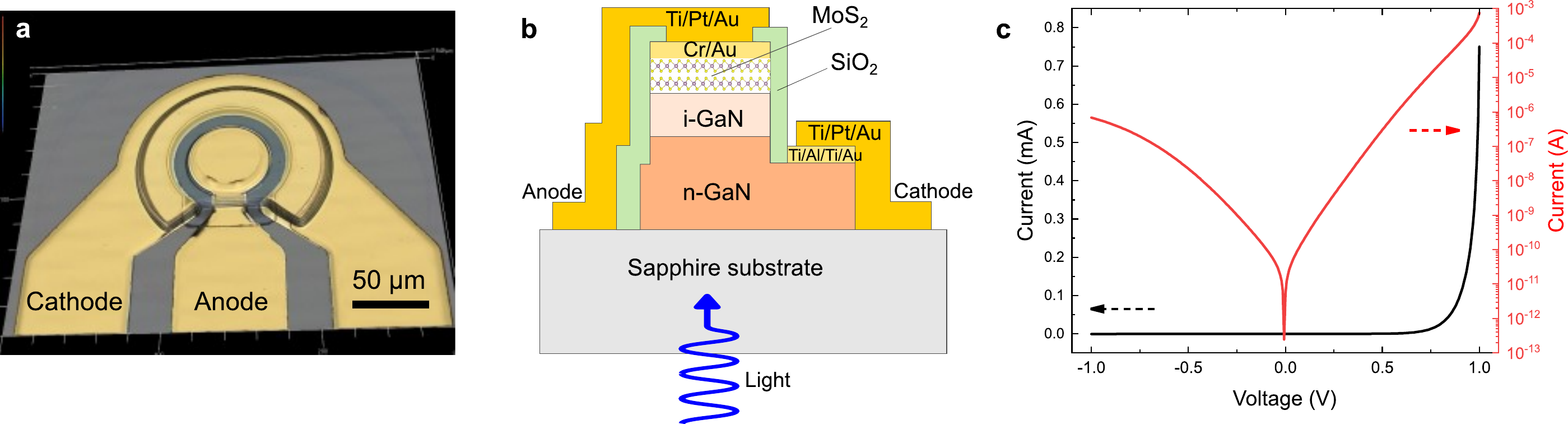}
    \caption{Device structure and electrical characteristics of MoS\(_2\)/GaN UTC-PD.
    \textbf{a}, Three-dimensional image of the fabricated device acquired using a laser microscope.
    \textbf{b}, Cross-sectional schematic image. The blue arrow illustrates backside illumination in our characterization.
    The device structure consists of three main layers: an n-GaN layer (thickness: 3.6 $\mu$m, carrier density: $4.22 \times 10^{19}$ cm$^{-3}$), an i-GaN layer (thickness: 0.3 $\mu$m, carrier density: $10^{15} \sim 10^{16}$ cm$^{-3}$), and a MoS\(_2\) layer (2--4 layers, sheet carrier density: $8.5 \times 10^{13}$ cm$^{-2}$).  
    \textbf{c}, I--V characteristics of the device measured under dark condition.}
    \label{fig:Device_Structure}
\end{figure}

Figure~\ref{fig:Device_Structure}\textbf{a} provides a three-dimensional image of the fabricated device, whose photosensitive area is 60 $\mu$m in diameter.
The device is designed to operate with illumination from the substrate side, as indicated in Fig.~\ref{fig:Device_Structure}\textbf{b}.
Photolithography and reactive ion etching (RIE) processes are utilized to form the mesa pattern of the device. 
The stacking structure comprises an n-GaN contact layer, an i-GaN carrier transport layer with the thickness of $0.3~\mu$m and the carrier density $\sim 10^{16}~$cm$^{-3}$, MoS$_2$ photoabsorption layers.
A passivation layer of SiO\(_2\) is deposited around the mesa structure. 
Subsequently, Cr/Au is deposited as the anode electrode, and Ti/Al/Ti/Au is deposited as the cathode electrode using sputtering techniques. 
Finally, Ti/Pt/Au is deposited as the pad electrode.

I--V characteristics of the device measured under dark conditions clearly exhibit diode-like rectifying behavior as in Fig.~\ref{fig:Device_Structure}\textbf{c}.
The summary of Hall effect measurements, sheet resistivities, mobilities, and carrier densities of MoS\(_2\) and GaN layers, are found in Supplementary Table~1.
Notably, the Hall measurements of the MoS\(_2\) layers indicate n-type conduction, which may be attributed to the sulfur atom vacancies \cite{Liu2013,Su2015}.
Transmission line model (TLM) measurements (described in Supplementary Fig. 1) confirms the ohmic contact between the anode electrode and MoS\(_2\) layers, indicating the rectification originates from the MoS\(_2\)/GaN heterojunction rather than Schottky barrier at the contact.
The physics behind the rectifying behavior of the device is discussed in Supplementary Section~2, where the two factors are considered: the crystal-momentum mismatch between the conduction electrons in MoS\(_2\) and GaN \cite{Kummel2017}, and the polarization charge of GaN \cite{Li_Zhang_Zhao_Yan_Liu_Wang_Li_Wei_2020}.

\section*{Experimental Verification} \label{sec:4}

\begin{figure}[htbp]
    \centering
\includegraphics[width=0.5\textwidth]{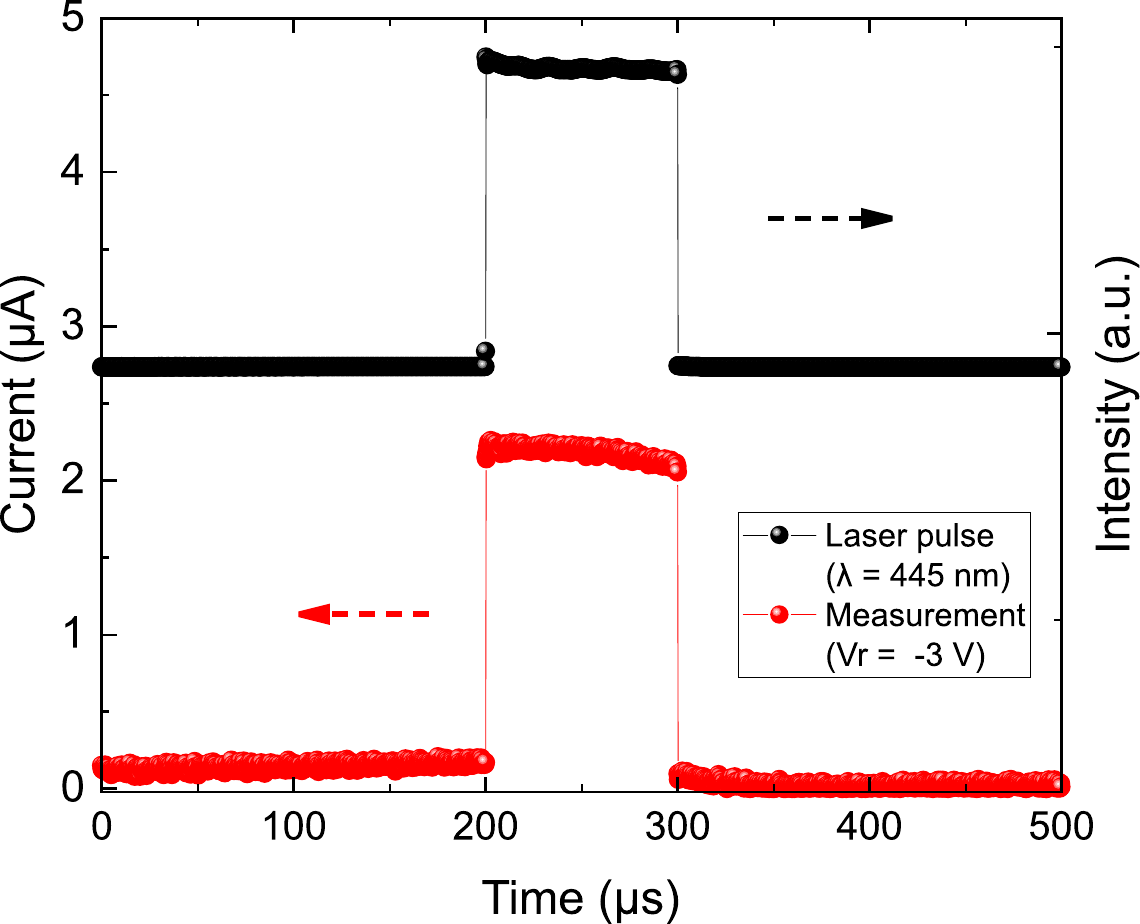}
\caption{Photocurrent signal of the device is shown as the red line, while the pulse waveform signal of the incident light is shown as the black line. 
Here, the latter signal is taken by a Si PIN-photodiode (Hamamatsu Photonics; S9055) as a reference. 
The rise and fall times of the incident light are on the order of microseconds, which closely follows the pulse waveform of the incident light.}
\label{fig:Photocurrent_Response}
\end{figure}

We evaluate the large-signal response of the device by measuring the time-domain photocurrent signal when illuminating the light with a wavelength of 445~nm and a pulse width of 100 $\mu$s. 
In Fig.~\ref{fig:Photocurrent_Response}, the black line represents the pulse waveform of the incident light, which is taken by a Si PIN-photodiode (Hamamatsu Photonics; S9055), as a reference while the red line represents the photocurrent signal of the device. 
The photocurrent signal closely follows the pulse waveform of the incident light, where the rise and fall times are on the order of microseconds which are limited by the time-response of the light source itself.
Note that the slight distortions of baseline and the top of the pulse in the photocurrent signal are mainly due to the peripheral circuits.
The photocurrent response is thus three to six orders of magnitude faster than previously reported MoS\(_2\)/GaN photodetectors~\cite{Wu2019,PulsedLaserDeposition2023,Moun2018,SelfPowered2023,Veeralingam2023}.
Such improvements can be attributed to the high-quality epitaxial MoS\(_2\)/GaN van der Waals heterointerface, which effectively reduces carrier trapping effects at the heterojunction.

\begin{figure}[htbp]
    \centering
\includegraphics[width=\textwidth]{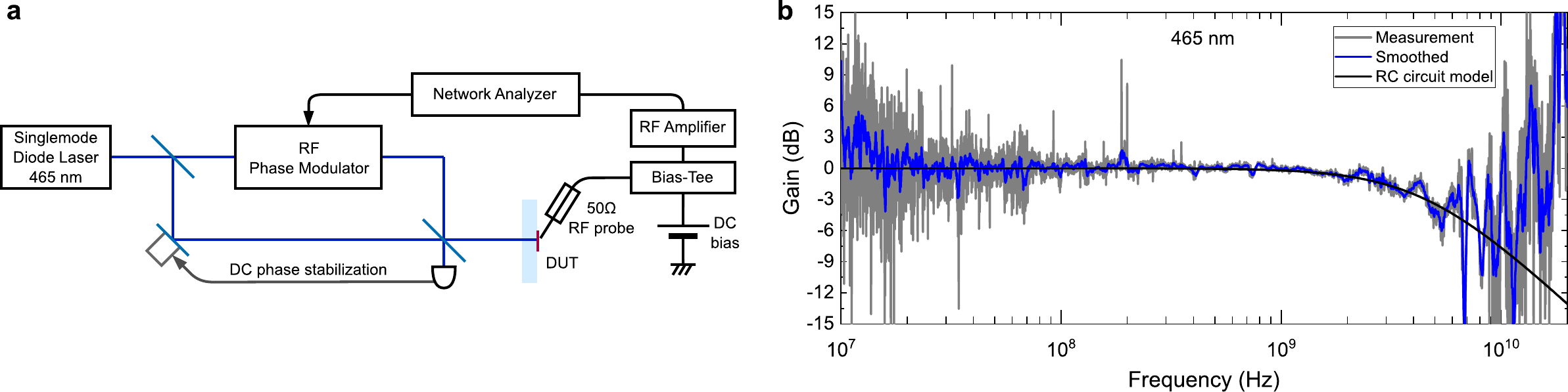}
    \caption{Frequency response of the device.
    \textbf{a}, Experimental setup for measuring the frequency response of the fabricated MoS\(_2\)/GaN UTC-PD.
    A single-mode laser diode (LD) operating at 465 nm is modulated by a waveguide phase modulator to generate small intensity modulation via a stabilized Mach--Zehnder interferometer.
    An AC photocurrent signal is fed into a vector network analyzer through a  transmission line probe and a wideband amplifier matched at 50 $\Omega$.
    \textbf{b}, Frequency response characteristics of the device at a wavelength of 465 nm with a reverse bias of 3 V, normalized by low-frequency sensitivity.
    Non-flat frequency response of the measurement system is calibrated by a reference GaAs PIN-photodiode that have a flat frequency response up to 20 GHz.
    The raw signal (gray line) contains large fluctuations both in low and high frequencies due to the 1/f noise and the crosstalk of the radiation from the driving signal.
    Smoothing is applied (blue line) to see the salient feature of the frequency response.
    A simple RC low-pass filter response (black line) is also shown.}
    \label{fig:FreqResp_Setup}
\end{figure}

To investigate the characteristics of the device in much higher frequency regions, we examine the frequency response of the device with intensity-modulated light (Fig.~\ref{fig:FreqResp_Setup}\textbf{a}).
Figure~\ref{fig:FreqResp_Setup}\textbf{b} demonstrates that the MoS\(_2\)/GaN PD achieves a $-$3~dB bandwidth of approximately 5~GHz at a reverse bias of 3 V.
This response is well explained by an RC low-pass filter model, where R corresponds to the external impedance of $\mathrm{R}_\mathrm{load}$ = 50 $\Omega$, and C corresponds to the i-GaN depletion capacitance of $\mathrm{C}_\mathrm{PD}$ = 0.8 pF.
The capacitance is identified by capacitance-voltage (C--V) measurement as described in Supplementary~Fig. 2.
The cutoff frequency of 5~GHz is thus neither due to carrier escape delay in the MoS\(_2\) layer nor carrier transit time in the i-GaN layer.
We can anticipate a faster response by minimizing the capacitance by, for instance, optimizing the thickness of the electron transport layer, and/or miniaturizing the size of the photosensitive area.

\begin{figure}[htbp]
    \centering
\includegraphics[width=\textwidth]{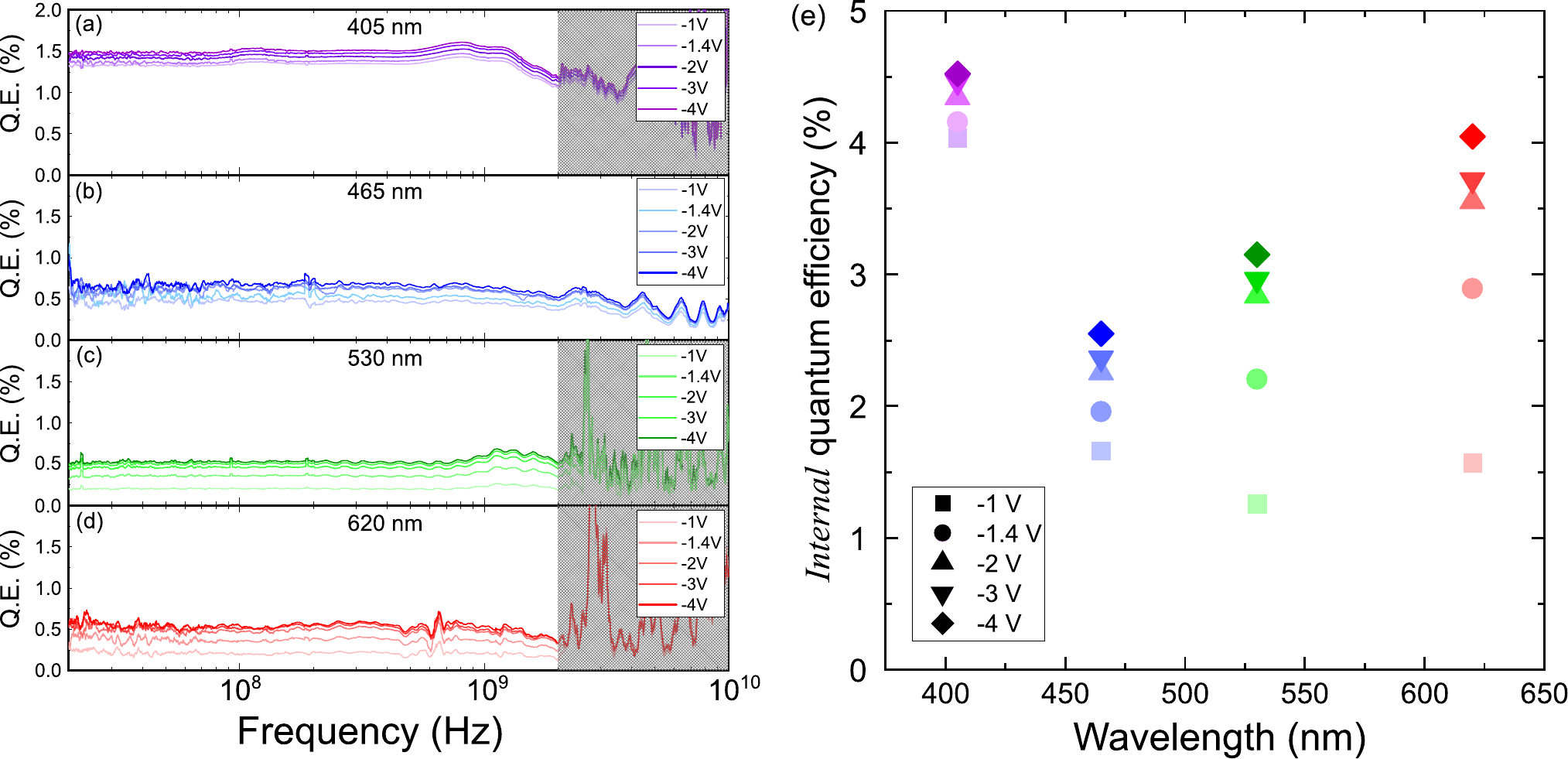}
    \caption{\textbf{a}-\textbf{d}, Wavelength dependence of the frequency response, measured using laser diodes (LD) at 405~nm (\textbf{a}), 465~nm (\textbf{b}), 530~nm (\textbf{c}), and 620~nm (\textbf{d}). 
    Each response is represented in terms of the quantum efficiency (Q.E.) calibrated by a commercially available Si PIN-PD, whose quantum efficiency is specified by the manufacturer.
    We only show the results with smoothing to eliminate the low and high-frequency noise.
    In 6\textbf{a}, 6\textbf{c}, 6\textbf{d}, frequency region above 2 GHz is grayed-out to indicate the signal is dominated by crosstalk.
    The lines in each figure correspond to different reverse bias voltage applied to the device. 
    \textbf{e}, \textit{Internal} quantum efficiencies for varying wavelengths and reverse biases.
    The \textit{internal} quantum efficiency is defined as the quantum efficiency (depicted in 6\textbf{a}-6\textbf{d}) divided by the absorption (depicted in Fig.~\ref{fig:MoS${}_2$GaN_overview}\textbf{d}).}
    \label{fig:Wavelength_FreqResp}
\end{figure}

Next, we evaluate the dependence of sensitivity and frequency response on wavelength and reverse bias voltage.
Figures~\ref{fig:Wavelength_FreqResp}~\textbf{a}-\textbf{d} show the results for the wavelengths of the light at 405~nm, 465~nm, 530~nm, and 620~nm, respectively. 
For measurements at 405~nm, 530~nm, and 620~nm, the intensity modulation is performed by direct modulation of the laser diodes rather than by using the separate phase modulator described in Fig.~\ref{fig:FreqResp_Setup}\textbf{a}.
Even after smoothing, nonignorable noise remains above 2~GHz (shaded region) for those results, which is mainly due to the larger electrical crosstalk caused by the direct modulation circuit. 
The broad sensitivity across the visible wavelength suggests that electron-hole pair generation occurs in the MoS$_2$ layer rather than GaN layers.
Thus, the photocurrent signal corresponds to electron transit in the i-GaN layer, in accordance with the UTC-PD model shown in Fig.~\ref{fig:MoS${}_2$GaN_overview}\textbf{a}.
Highest quantum efficiency is obtained at 405 nm, reaching approximately 1.5\%, while the efficiencies are lower around 0.5\% for other wavelengths (465~nm, 530~nm, and 620~nm), which is consistent with the photoabsorption spectrum (Fig.~\ref{fig:MoS${}_2$GaN_overview}\textbf{d}).

Figure~\ref{fig:Wavelength_FreqResp}\textbf{e} shows the \textit{internal} quantum efficiencies for varying wavelengths and reverse biases. 
Here, the \textit{internal} quantum efficiency is defined as the quantum efficiency divided by the absorption depicted in Fig.~\ref{fig:MoS${}_2$GaN_overview}\textbf{d}, expressing what portion of electrons generated in the MoS$_2$ layer turns into the photocurrent.
The fabricated devices exhibit relatively low \textit{internal} quantum efficiencies, ranging approximately from 1\% to 5\%, as shown in Fig.~\ref{fig:Wavelength_FreqResp}\textbf{e}.  
Note that, as the reverse bias increases, the \textit{internal} quantum efficiency also increases for all wavelengths. 
Furthermore, the reverse bias dependence of the \textit{internal} quantum efficiency is more significant for longer wavelengths.
The plausible explanation for these peculiar behaviors of the \textit{internal} quantum efficiencies with respect to wavelengths and reverse biases is suggested in Supplementary Section~3.
We claim that the momentum-space alignment between absorption and transport bands should become an important guideline in the design of high-performance 2D-material/GaN photodetectors, which has largely been overlooked in the literature.

\section*{Conclusion} \label{sec:6}

We successfully fabricated an UTC-PD based on a MoS$_2$/GaN van der Waals heterojunction, where the type-II aligned MoS\(_2\) layer selectively inject photoelectrons to the i-GaN transport layer.
The fabricated UTC-PD worked at entire visible wavelengths with the quantum efficiency on the order of 1\%.
Our device exhibits the excellent frequency response, which corresponds to three to six orders of magnitude faster than those reported in earlier studies, thanks to the high-quality epitaxial van der Waals heterointerface between MoS\(_2\)/GaN.

Through this study, we proved the feasibility of UTC-PDs for visible light by incorporating a 2D semiconductor layer as a photoabsorption layer and a wide-bandgap GaN film as an electron transport layer. 
Further material exploration and structural optimization are expected to provide additional breakthroughs, opening new paradigms in the field of high-speed visible optoelectronics.

\section*{Method} \label{sec:7}

\subsection*{Growth of MoS\(_2\)}
The MoS\(_2\) layers were directly grown onto a GaN film on a c-plane sapphire substrate using MOCVD system
(Oxford Instruments; Nanofab 1000 Agile).
The precursors used for the MoS\(_2\) growth are molybdenum hexacarbonyl, Mo(CO)\(_6\) (Japan Advanced Chemicals) for metal, and hydrogen sulfide, H\(_2\)S for chalcogen.
The growth is conducted under the condition in which the pressure is 4 Torr and the temperature is 800$^\circ$C.

\subsection*{STEM analysis}
To obtain atomically resolved information on the MoS\(_2\)/GaN van der Waals heterojunction, a STEM analysis has been performed as shown in Fig. 1\textbf{c} and Extended Data Fig. 1.
The sample was prepared by thinning using a micro-sampling method with a Focused Ion Beam (FIB).
The observations were made along the a-plane direction [11\(\bar{2}0\)] of GaN using STEM (JEOL; JEM-ARM200F).

\subsection*{Fabrication of the device}
Photolithography and reactive ion etching (RIE) process were used to make the mesa patterns of the device. 
The photosensitive area is 60 $\mu$m in diameter.
The anode electrode is composed of Cr/Au (10~nm/200~nm) while the cathode consists of Ti/Al/Ti/Au (10 nm/300 nm/150 nm/50 nm) deposited by sputtering.

\subsection*{Experimental setup for the frequency response measurement}
A single-mode laser diode (LD) operating at 465 nm (NICHIA; NDB4916) is modulated by a waveguide phase modulator (AdvR; WPM-K0450) to generate small intensity modulation via a stabilized Mach--Zehnder interferometer (Fig.~\ref{fig:FreqResp_Setup}\textbf{a}).
An AC photocurrent signal is fed into a vector network analyzer (ROHDE \& SCHWARZ; ZNA67) through a  transmission line probe (TECHNOPROBE; TP26-GSG-350-A) and a wideband amplifier matched at 50 $\Omega$.
Non-flat frequency response of the measurement system is calibrated by a commercially available GaAs PIN-photodiode (TRUMPF; ULMPIN-25-TT).
A wavelength dependence of the frequency response, measured using LD at 405~nm (NICHIA; NDV4316), 465~nm (NICHIA; NDB4916), 530~nm (NICHIA; NDG4716), and 620~nm (Hangzhou BrandNew Technology; custom-made), as shown in Fig.~\ref{fig:Wavelength_FreqResp}. 
Q.E. is calibrated by a commercially available Si PIN-photodiode (Hamamatsu Photonics; S9055).
The pulsed light, as shown in Fig. 4, was generated by directly modulating a 445 nm LD to a pulse width of 100 $\mu$s using a variable pulse width driver (PicoLAS GmbH; LDP-V 03-100).

\subsection*{DFT calculations}
Band structure of bilayer MoS\(_2\) (2H phase), as shown in Supplementary Fig. 4\textbf{a}, was calculated using density functional theory (DFT)
All calculations were performed using the Vienna Ab initio Simulation Package (VASP).
The local density approximation of Ceperley and Alder (CA) was used for the exchange-correlation functional. 
The projector augmented-wave (PAW) potentials were used to describe the core-valence interaction. 
A plane-wave energy cutoff was set to 600~eV and an energy convergence criteria of 10\(^{-7}\) eV was used. 
The atomic positions were relaxed until the residual forces on all atoms became less than $5 \times 10^{-3}$ eV/$\text{\AA}$. 
For the initial structure of the bilayer MoS\(_2\), we adopted a- and b-axes lattice constants and atomic positions of the bulk 2H-MoS\(_2\), 
and the vacuum layer of about 25 $\text{\AA}$.
The atomic positions of bilayer MoS\(_2\) model were relaxed, while the cell parameters were fixed. 
For the $K$-point sampling of Brillouin-zone integrals, we used a $8 \times 8 \times 2$ for the bulk 2H-MoS\(_2\), 
and a $8 \times 8 \times 1$ for the bilayer MoS\(_2\).

\subsection*{Characterisations}
The 3D image of the device (Fig.~\ref{fig:Device_Structure}\textbf{a}) was acquired using a laser microscope (Keyence; VK-X3000).
The Raman spectra (Fig.~\ref{fig:MoS${}_2$GaN_overview}\textbf{b}) were measured with Raman spectrometer (Renishaw; inVia reflex) using 532 nm excitation.
The absorption spectra (Fig.~\ref{fig:MoS${}_2$GaN_overview}\textbf{d}) were measured by a spectrophotometer (Hitachi High-Tech; U-4100).
The atomic resolution cross-sectional STEM images (Fig.~\ref{fig:MoS${}_2$GaN_overview}\textbf{c}) were measured using STEM (JEOL; JEM-ARM200F). 
The I--V characteristics of the device (Fig.~\ref{fig:Device_Structure}\textbf{c}) measured using a device analyzer (Keysight; B1505A).
The surface potential of the materials was measured using a KPFM (SHIMADZU; SPM-9700HT).

\section*{Data availability}
The data that support the findings of this study are available from the corresponding author upon reasonable request.

\section*{Acknowledgments}
We thank T. Kishino for the GaN epitaxially growth on Sapphire substrate, Y. Nakagawa for the help with MoS\(_2\) growth, M.~Sano, K.~Omae, S.~Nagahama, T.~Mukai, and Y.~Narukawa for the fruitful discussions.

\section*{Author contributions}
T.K. and T.S. designed the UTC-PD based on a MoS\(_2\)/GaN heterojunction. T.K. prepared and analysed the MoS\(_2\) layers and fabricated the device.
T.S. designed the experimental setup for the frequency response measurement. T.K. characterized the device. A.I. performed the DFT calculations.
T.K., T.S., K.U., and H.O. wrote the paper. H.O. organized all the efforts on this work. 
All authors contiributed to the discussion concerning the experimental results.

\section*{Competing interests}
The authors declare no competing interests.

\newpage

\begin{figure}[htbp]
    \centering
    \includegraphics[width=\textwidth]{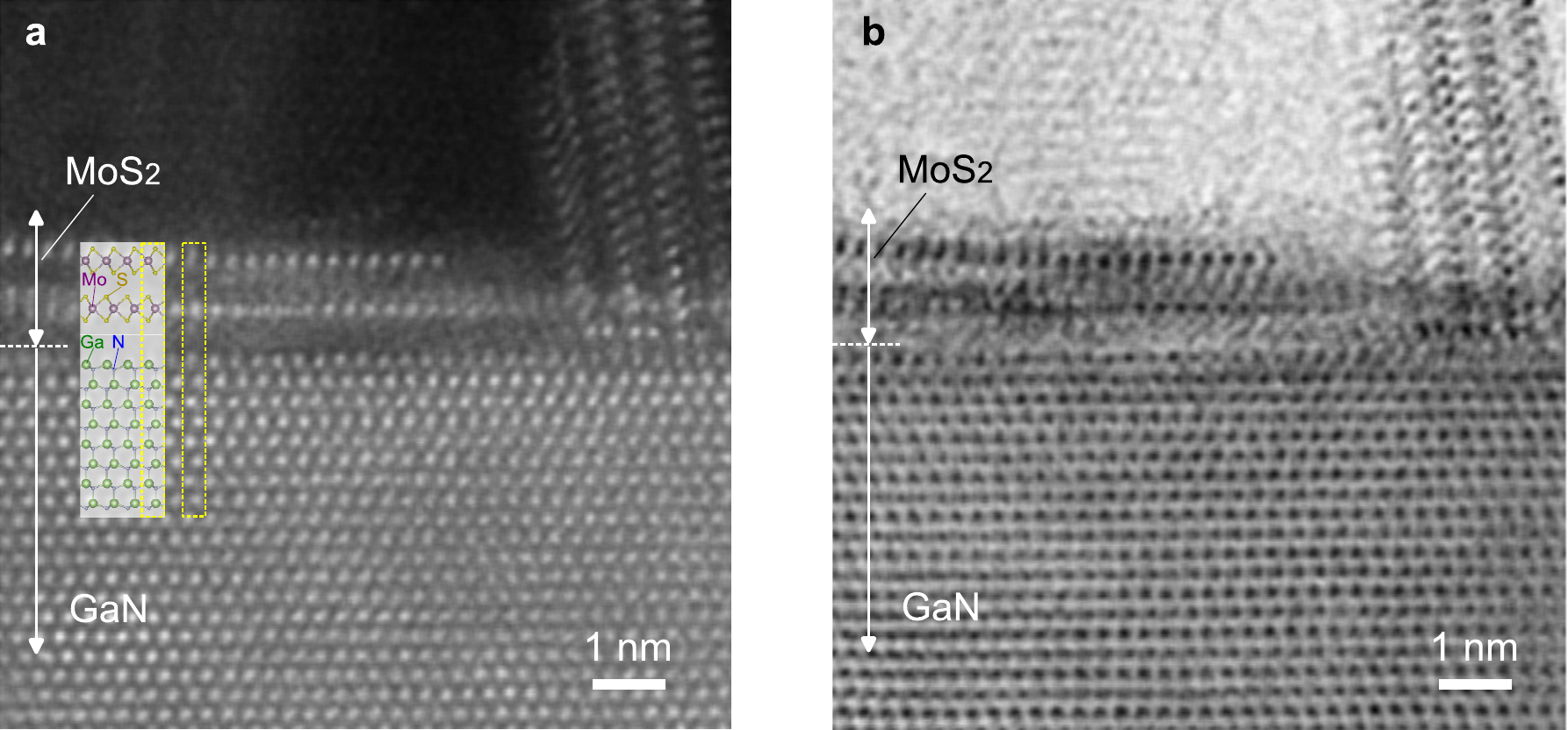}
    \captionsetup{labelformat=empty}
    \caption{Extended Data Fig. 1: Atomic resolution cross-sectional STEM images of MoS\(_2\)/GaN van der Waals heterostructure. \textbf{a}, High-angle-annular-dark-field (HAADF) mode. \textbf{b}, Annular bright field (ABF) mode. (\textbf{a}) is a STEM image observed in HAADF mode, and it is the same as that shown in Fig. 1\textbf{c} in the main text. 
    In the HAADF mode, obtained by collecting electrons scattered at large angles, the bright contrast is associated to the columns of atoms (i.e., Mo, Ga) with high atomic number.
    Additionally, a schematic diagram of the atomic arrangement of MoS\(_2\) and GaN is superimposed on the STEM image. 
    As indicated in the yellow dotted line, it can be seen that the MoS\(_2\) layers are epitaxially grown on the GaN film.
    On the other hand, the STEM image in ABF mode shown in (\textbf{b}) allows observation of atoms with high and lower atomic numbers (i.e., S, N) as dark spots with higher and lower intensities, respectively.}
    \label{fig:STEM}
\end{figure}

\newpage

{\centering \LARGE Supplementary Information \\[1em] Uni-Traveling-Carrier Photodiode Based on MoS\(_2\)/GaN van der Waals Heterojunction for High-Speed Visible-Light Detection}

\tableofcontents

\newpage

\section{Electrical characteristics} 

Supplementary Fig.~1\textbf{a} and \textbf{b} show the current-voltage (I--V) curves acquired on n-GaN layer and MoS\(_2\) layer of Transmission Line Model (TLM) structures, respectively.
Additionally, schematic diagrams of the measurements are presented in the inset of the figures.
The plots in the figure represent the distances between the electrode pads of 10 $\mu$m, 15 $\mu$m, and 20 $\mu$m, respectively.
The results indicate n-type ohmic contacts formed between the MoS\(_2\) layer and the anode electrode (Cr/Au), as well as between the GaN layer and the cathode electrode (Ti/Al/Ti/Au), with a linear I--V curve.

\begin{figure}[htbp]
    \centering
\includegraphics[width=\textwidth]{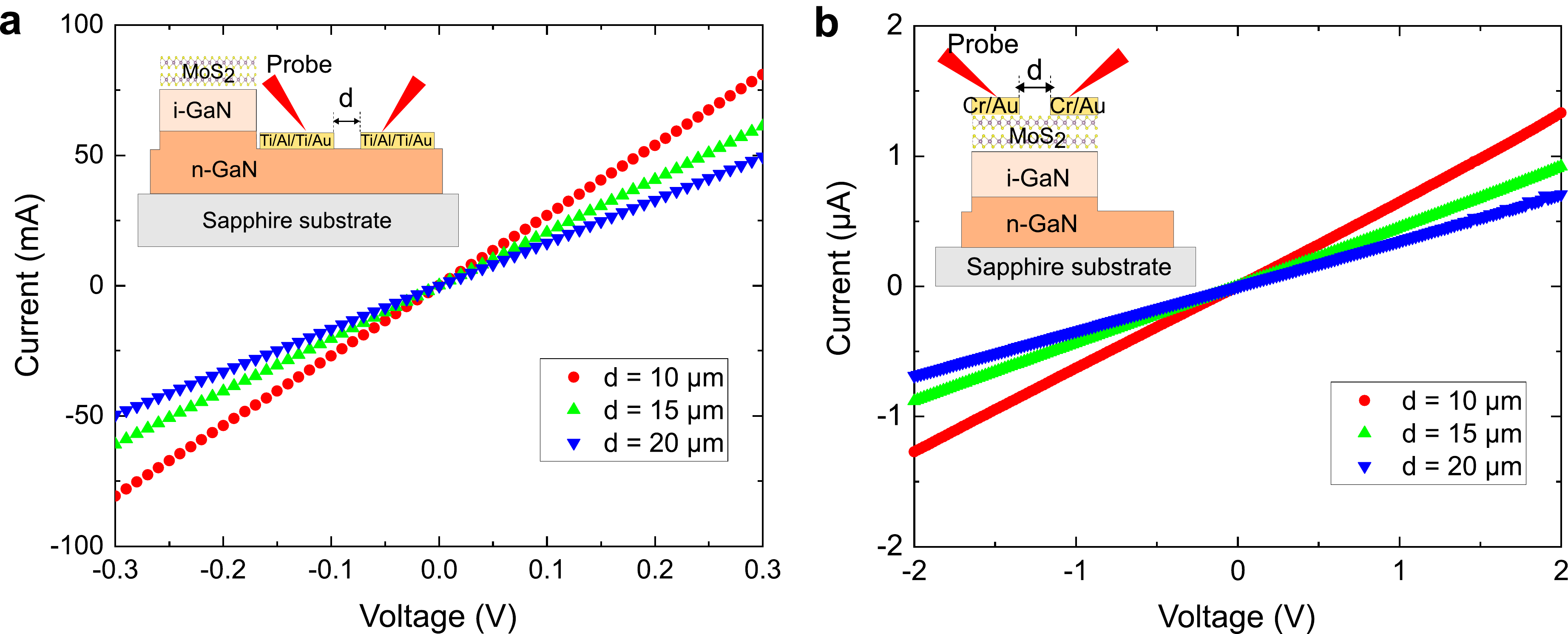}
\captionsetup{labelformat=empty}
    \caption{Supplementary Fig. 1: Current-voltage (I--V) curves acquired on pads of Transmission Line Model (TLM) structures at a distance of 10 $\mu$m, 15 $\mu$m, 20 $\mu$m. \textbf{a}, I--V curve on n-GaN layer. \textbf{b}, I--V curve on MoS\(_2\) layer. The insets show the schematic diagrams of the measurements.}
    \label{fig:TLM_IV}
\end{figure}

\newpage

Supplementary Fig.~2 shows the capacitance-voltage (C--V) characteristics of the MoS\(_2\)/GaN van der Waals heterojunction. The C--V characteristic was measured with bias applied to contacts on MoS\(_2\) and was consistent with that of a reverse-biased p-n junction, as shown in the inset of the figure.
It is noteworthy that since the i-GaN layer is fully depleted, the capacitance of the device remains almost unchanged even with an increase in the reverse bias voltage.

\begin{figure}[htbp]
    \centering
\includegraphics[width=0.5\textwidth]{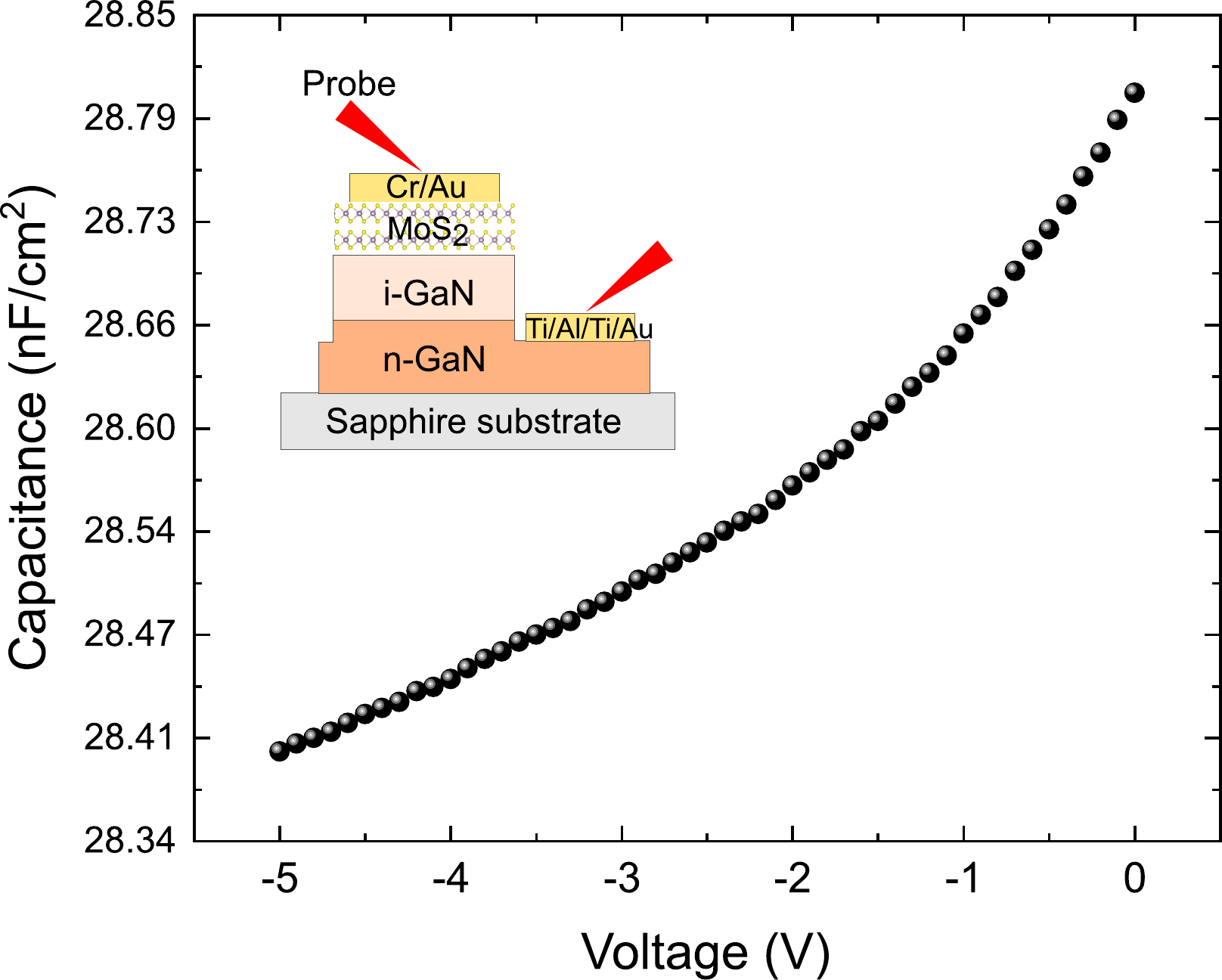}
\captionsetup{labelformat=empty}
    \caption{Supplementary Fig. 2: Capacitance-voltage (C--V) characteristics of the MoS\(_2\)/GaN van der Waals heterojunction. The inset shows the schematic diagrams of the measurement.}
    \label{fig:CV_curve}
\end{figure}

\newpage

The carrier concentration and doping type of the GaN film and the MoS\(_2\) layers was estimated on the basis of the results of the Hall effect measurement, as shown in Supplementary Table~1. 
This measurement was carried out using a Hall effect measurement system (Nanometrics; HL5500) under the conditions of room temperature, an AC current of 0.1 $\mu$A, and a magnetic field of 0.5 T, utilizing the Van der Pauw method.
MoS\(_2\) exhibited n-type semiconductor characteristics due to the vacancies of sulfur atoms, as shown in the main text. 

\begin{table}[htbp]
\centering
\captionsetup{labelformat=empty}
\caption{Supplementary Table 1: The results of room-temperature Hall effect measurements for GaN and MoS\(_2\)}
\begin{tabular}{ccccc}
\hline \hline
Sample & Sheet Resistivity ($\Omega$/$\square$) & Mobility (cm$^2$/V$\cdot$s) & Carrier Density & Type Conductivity \\ \hline
GaN & $10.8$ & $137$ & $4.22 \times 10^{19}$ cm$^{-3}$ & N \\
MoS\(_2\) & $3.7 \times 10^3$ & $19.5$ & $8.5 \times 10^{13}$ cm$^{-2}$ & N \\ \hline \hline
\end{tabular}
\label{tab:Hall}
\end{table}

\newpage

\section{Discussion of Rectification characteristics}

We discuss the rectifying behavior of the device shown in Fig.~3\textbf{c}.
According to the Hall effect measurement (Supplementary~Table~1) and the band alignment measured by KPFM (shown in Fig.~2), the heterojunction between MoS\(_2\) and GaN of our device is isotype and there is only a small energy difference in the conduction band.
To explain the observed rectification in the I--V characteristics, we introduce two factors; the crystal-momentum mismatch of the conduction electrons in MoS\(_2\) and GaN \cite{Kummel2017}, and polarization charge of GaN \cite{Li_Zhang_Zhao_Yan_Liu_Wang_Li_Wei_2020}.

Supplementary Fig. 3\textbf{a} (Supplementary Fig. 3\textbf{b}) shows the band diagram of the MoS\(_2\)/GaN van der Waals heterojunction under forward (reverse) bias.
We highlight the bands at $\Gamma$-point (solid lines) and $K$-point (dashed lines) in the Brillouin zone, which are depicted separately to illustrate the effect of crystal-momentum mismatch of the conducting electrons at the heterointerface.
The conduction band minimum of GaN is at $\Gamma$-point, while that of MoS\(_2\) is at $K$-point.
Under the forward bias, electrons injected from the cathode contact reside at conduction band minimum of GaN i.e., at $\Gamma$-point.
These electrons can flow into the conduction band at the $\Gamma$-point of MoS\(_2\) without appreciably changing their crystal momentum as in Supplementary Fig. 3\textbf{a}.
After flowing into the the conduction band at the $\Gamma$-point of MoS\(_2\), they rapidly decay into the conduction band at $K$-point of MoS\(_2\) and contribute to forward current.
Under the reverse bias, on the other hand, electrons injected from the anode contact resides at the conduction band minimum at $K$-point of MoS\(_2\).
They can neither flow into the conduction band at the $K$-point of GaN due to the large energy barrier, nor into the conduction band at the $\Gamma$-point of GaN due to the large crystal-momentum mismatch as shown in Supplementary Fig. 3\textbf{b}.

Spontaneous polarization in GaN can also introduce an additional potential barrier.
Specifically, interfacing the nonpolar MoS\(_2\) layer with polar GaN generates negative polarization charges at the GaN/MoS\(_2\) interface. 
These charges are enhanced by compressive strain in the GaN-on-sapphire structure through piezoelectric effects. 
The polarization-induced charges lead to electron accumulation at the MoS\(_2\)surface and cause depletion in the adjacent i-GaN layer even without external bias. 
We highlight this space-charge distribution as the slope of the band diagrams schematically depicted in Supplementary Fig. 3.

\newpage

\begin{figure}[htbp]
    \centering
\includegraphics[width=\textwidth]{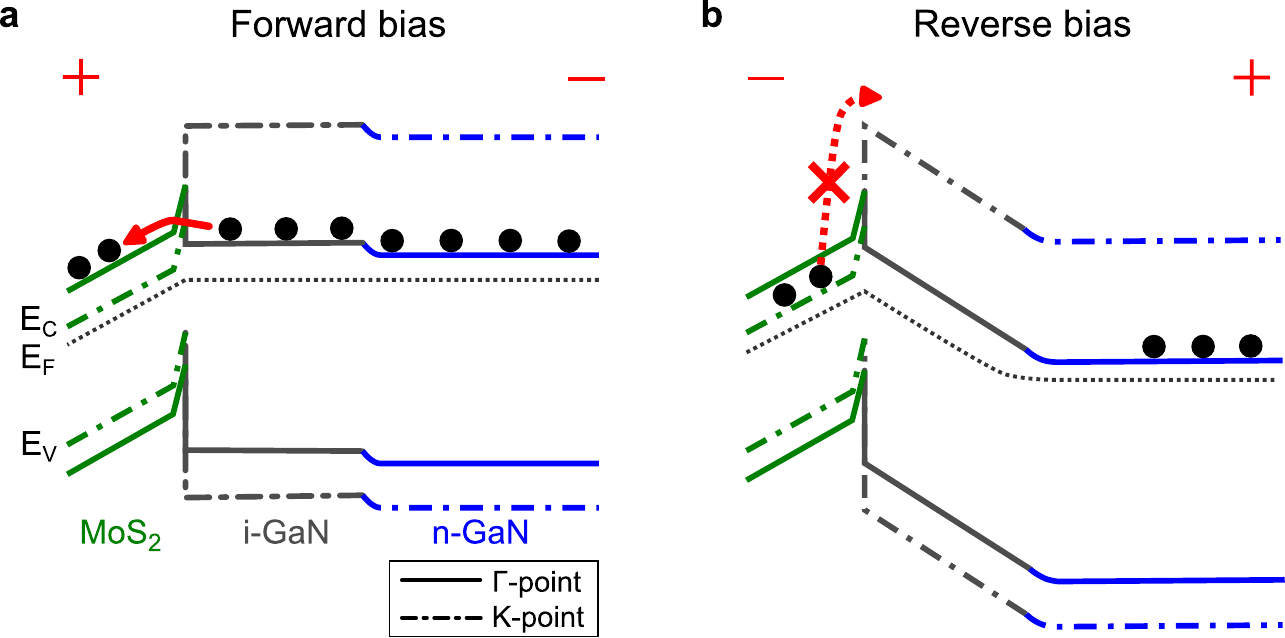}
\captionsetup{labelformat=empty}
    \caption{Supplementary Fig. 3: Band diagram of the MoS\(_2\)/GaN van der Waals heterojunction. \textbf{a}, Under the forward bias. \textbf{b}, Under the reverse bias. The solid lines represent the bands at $\Gamma$-point in the Brillouin zone, while the dashed lines represent the bands at $K$-point.}
    \label{fig:Band_Diagram}
\end{figure}

\newpage

\section{Discussion of the \textit{internal} quantum efficiencies}

The devices exhibit rather low \textit{internal} quantum efficiencies as shown in Fig.~6\textbf{e}. 
These values are significantly lower than those of conventional PN photodiodes, typically approaching near 100\%. 
The imperfect \textit{internal} quantum efficiencies indicate that certain mechanisms prevent a fraction of photogenerated carriers from contributing to the photocurrent, and these mechanisms may vary with wavelength and applied bias.

Supplementary Fig. 4\textbf{a} shows the band structure of bilayer MoS\(_2\) (2H phase) calculated using density functional theory (DFT), which models the electron-hole pair generation in our devices.
We focus on the C-exciton transition ($\sim$ 3 eV) lies on the $\Gamma$-point and the B-exciton transition($\sim$ 2 eV) on the $K$-point, indicated by the purple and red arrows in Supplementary Fig. 4\textbf{a}.
Each of these transitions can be excited by light with wavelengths of 405 nm and 620 nm, respectively.        
The simplified bands relevant to these two transitions are shown in Supplementary Fig. 4\textbf{b}, where, again, the solid lines represent the bands at $\Gamma$-point and the dashed lines represent the bands at $K$-point. 

For the B-exciton transition, photo-electrons are generated around the $K$-point of MoS\(_2\) and they cannot transfer to either the $K$-point or $\Gamma$-point of GaN, due to the large energy barrier or the large crystal-momentum mismatch as shown in Supplementary Fig. 4\textbf{b}.
These electrons eventually recombine within the MoS\(_2\) layer without contributing to the photocurrent.
With the reverse bias and the resultant band tilting, however, these electrons can be transferred to the $\Gamma$-point of GaN through thermally assisted tunneling~\cite{MahaveerSathaiya_Karmalkar_2006}, leading to the improved \textit{internal} quantum efficiency as shown in Fig.~6\textbf{e}. 

For the C-exciton transition, photo-electrons are generated around $\Gamma$-point of MoS\(_2\). 
Most of them decay to the conduction band minimum at the $K$-point of MoS\(_2\), and thus behave like those of B-exciton transition. 
However, some portion of the generated electrons can flow directly into the conduction band minimum of GaN without suffering from the crystal-momentum mismatch as shown in Supplementary Fig. 4\textbf{b}, leading to the relatively higher \textit{internal} quantum efficiency at 405 nm even under the moderate reverse bias. 

For other transitions caused by the light with wavelengths of 465 nm and 530 nm, the situations are similar to the case of the C-exciton transition except that there are no direct paths into the $\Gamma$-point of GaN. 
The generated electrons in the conduction band of MoS\(_2\) cannot flow into the conduction band at the $\Gamma$-point for essentially the same reasons for the B-exciton transition.

In summary, some of the distinctive characteristics of the MoS\(_2\)/GaN UTC-PD need detailed considerations of the band structures of MoS\(_2\) and GaN including crystal-momentum mismatch. 
Further quantitative studies are required to fully understand the electron dynamics in the MoS\(_2\)/GaN van der Waals heterojunction and to improve the performance of the UTC-PDs. 
We claim that the momentum-space alignment between absorption and transport bands should become an important guideline in the design of high-performance 2D-material/GaN photodetectors, which has largely been overlooked in the literature.

\newpage

\begin{figure}[htbp]
    \centering
\includegraphics[width=\textwidth]{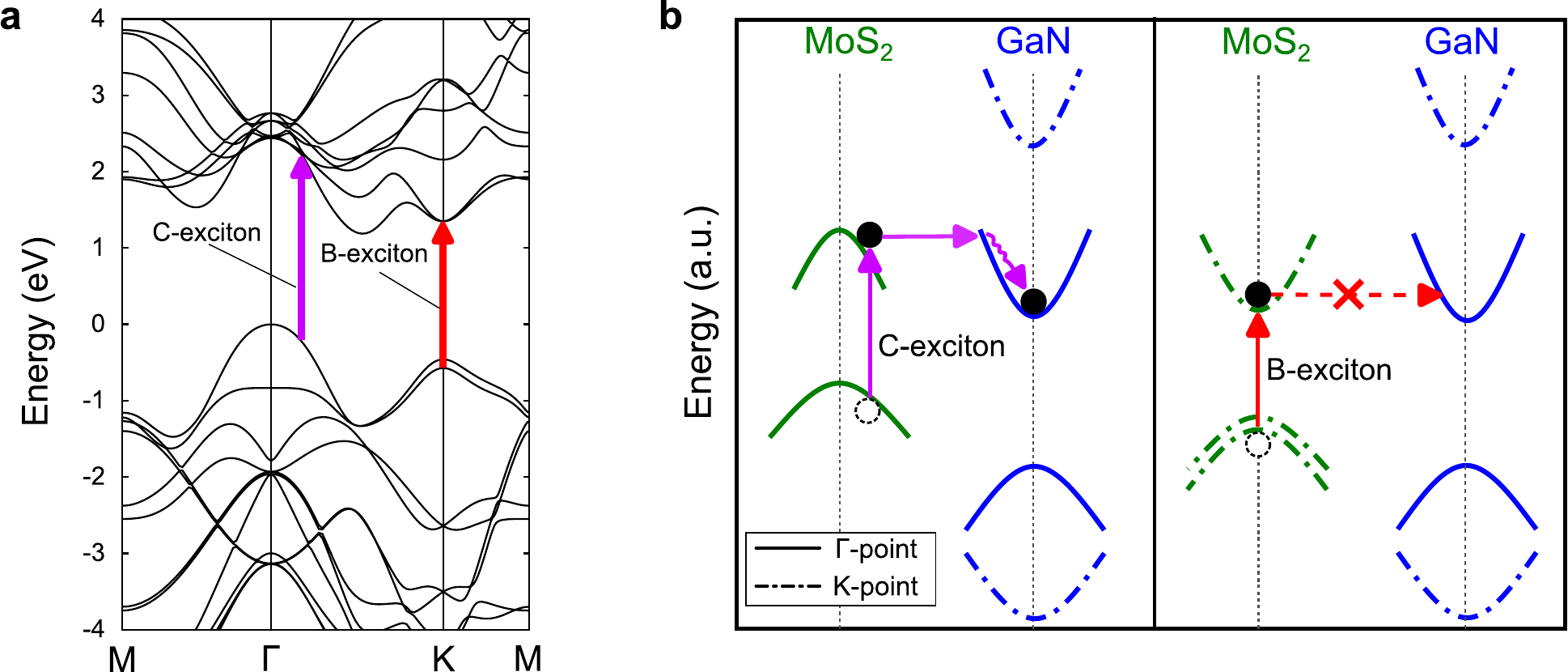}
\captionsetup{labelformat=empty}
    \caption{Supplementary Fig. 4: Band structure for the explanation of the electron dynamics in the MoS\(_2\)/GaN heterojunction. \textbf{a}, Band structure of bilayer MoS\(_2\) (2H phase), calculated using density functional theory (DFT).
    \textbf{b}, Simplified band diagrams relevant to the C-exciton and B-exciton transitions. The solid lines represent the bands at the $\Gamma$-point in the Brillouin zone, while the dashed lines represent the bands at the $K$-point.}
    \label{fig:Exciton_Bandstructure}
\end{figure}

\end{document}